# The Causal Role of Astrocytes in Slow-Wave Rhythmogenesis: A Computational Modelling Study

Leo Kozachkov, Konstantinos P. Michmizos

Laboratory for Computational Brain

Department of Computer Science, Rutgers University, Piscataway, NJ, USA

**Corresponding author**: Konstantinos P. Michmizos

Laboratory for Computational Brain

Department of Computer Science

Rutgers University

address: 110 Frelinghuysen Road, Piscataway, NJ, 08854

email: konstantinos.michmizos@cs.rutgers.edu

tel: 848-445-8841

fax: 732-445-0537

# 1. Abstract

Finding the origin of slow and infra-slow oscillations could reveal or explain brain mechanisms in health and disease. Here, we present a biophysically constrained computational model of a neural network where the inclusion of astrocytes introduced slow and infra-slow-oscillations. We show how astrocytes can modulate the fast network activity through their slow inter-cellular $Ca^{2+}$ wave speed and amplitude and possibly cause the oscillatory imbalances observed in diseases commonly known for such abnormalities, namely Alzheimer's disease, Parkinson's disease, epilepsy, depression and ischemic stroke. This work aims to increase our knowledge on how astrocytes and neurons synergize to affect brain function and dysfunction.

# 2. Introduction

Brain oscillations are periodic fluctuations in neuronal excitability that convey information in space and time [1, 2]. Slow oscillations (SOs; 0.1–1 Hz) and infra-slow oscillations (ISOs; 0.01–0.1 Hz) [3] characterize the idle "default" state of the brain [4] and orchestrate other default-state patterns such as delta-waves, sleep-spindles, and K-complexes [5, 6]. While the origin of these oscillations remains elusive [7], they have been associated with primitive brain functions, such as temperature regulation [8], hormone secretion [9] and memory consolidation [10], whereas their disturbance is often related to neurological disorders [11, 12]. Interestingly, SOs and ISOs share several distinctive patterns with the oscillatory activity exercised by the networks of the most abundant type of glial cells, the astrocytes [13, 14]: The neural activity recurs in a wave-like pattern with the same frequencies that astrocytes use to communicate with each other [15-17]; it can also be present even in the absence of stimulation, similarly to spontaneous inter-astrocyte communication [18-21].

Astrocytes, long believed to only give structural and nutritional support to neurons, are now regarded as key cells in neural information processing [22-26]. Astrocytes use processes, the long, thin tendrils protruding from their main cell bodies, to make contact with local neural synapses [26-29]. This three-body arrangement among pre- and post-synaptic neurons and the astrocyte process is named *tripartite synapse* [23, 26]. An individual astrocyte can form up to $10^5$ tripartite synapses [23, 26] and envelop up to 8 neuronal somata [30], with a single neuron belonging exclusively to a single astrocyte [31].

Contrary to neurons, which transmit information quickly by spikes, astrocytes are believed to use slowly propagating intercellular $Ca^{2+}$ waves [16, 18, 32-36]. $Ca^{2+}$ waves appear to recur at discrete origination sites with an infra-slow frequency [20, 33]. Although the mechanism by which intercellular $Ca^{2+}$ waves propagate is still unclear, both experimental and theoretical studies suggest that it involves a combination of direct, gap-junctional communication [32, 34] as well as extracellular, diffusion-driven communication [16, 37]. The average speed of astrocyte $Ca^{2+}$ wave varies considerably between studies [36], with values ranging from 8 µm/s [38] to 60 µm/s [37, 39]. Reported propagation distances for $Ca^{2+}$ waves vary between 4 astrocytes [16] and 100 astrocytes [36]. While the exact role of $Ca^{2+}$ waves remains unclear, even early studies showed that they can have both excitatory and inhibitory neuromodulatory effects [34, 35].

Astrocytes form dense syncytia, functional networks spanning across brain regions. Such regions include the retina [18, 34, 35], hippocampus [24, 27, 30, 40, 41], cortex [7, 42, 43], cerebellum [33] and thalamus [15]. Astrocyte networks tile the brain in a circumferential grid pattern [30] forming *synaptic* *islands*, which are non-overlapping domains of influence over nearby neurons [31]. In the synaptic island view of neural organization, "clusters of synapses are coordinated at nodes that correspond to the domain of individual astrocytes" [31]. However, to date, despite the multitude of studies that experimentally relate the astrocytic function with the slow oscillatory activity, there is still scarcity of studies, either experimental or theoretical, that have proposed a generation mechanism for these oscillations and therefore aim to understand how astrocytes modulate local neural activity in biological networks.

Being regarded as "a primary source for generating neural activity" [19], astrocytes have long been linked to SOs and ISOs through interacting with two synaptic sites [7, 41-43]. Specifically, astrocytes are found to modulate oscillations through co-activation of neuronal Adenosine A1 and N-methyl-D-aspartate (NMDA) receptors [24, 44]. However, up to now there is no suggested mechanism that connects the neural modulation to the long been observed slow oscillatory patterns of activity.

Toward this goal, we built a neurophysiologically plausible model of an astrocyte-neuron network that proposes a possible mechanistic explanation for the origination of SOs and ISOs as a result of astrocyte-neuron interactions during $Ca^{2+}$ wave events. By combining astrocytic pre-synaptic inhibition [22] with post-synaptic excitation [45], and replicating the connectome of astrocyte networks [30, 31], our proposed model faithfully reproduced the slow oscillations of neural activity, namely the ISO, derived from the $Ca^{2+}$ wave origination rate, and the SO, coming from the wave propagation speed. By varying the parameters of our model within their reported range, we also aimed to infer the possible key mechanisms for the oscillatory imbalances observed in case of diseases commonly known for exhibiting slow oscillatory abnormalities, namely Alzheimer's disease (AD), Parkinson's disease (PD), epilepsy, ischemic stroke and depression.

## 3. Methods

### *a. Network Architecture*

The astrocyte-neural network model consists of generic excitatory neurons organized in a feedforward fashion. The first layer of neurons served as the input layer to the network. The neurons at that layer were driven by a current drawn from a uniform distribution between 0 and 360 pA and exhibited an average firing rate of ~10 spikes per second throughout the duration of the simulation. In the subsequent layers, synaptic islands were formed between a single astrocyte and 4 neurons, to replicate the average number of neurons that create tripartite synapses with a single astrocyte [31]. We fixed the number of layers at 5, to replicate the approximate number of radially positioned astrocytes on which a single $Ca^{2+}$ wave has been reported to spread in *in-vitro* connectivity studies [30, 36]. The amplitude of the $Ca^{2+}$ wave decreased as it propagated through the 5 astrocytes by 7.9% per astrocyte. Each neuron was fully connected to the neurons of the next layer. Neurons in layers other than the input layer were additionally driven by a small noisy basal current, drawn from a uniform distribution between 0 and 180 pA to represent all the other inputs that a neuron receives from sources outside the network [15, 46]; this, in conjunction with astrocytic stimulation, resulted in them having average firing rates of firing rates of roughly [14, 20, 26, 32, 38] spikes/sec throughout the course of the simulation, for each layer, respectively [47].

The astrocytes were modelled as point cells positioned 50μm away from one another, in accordance with experimental data [30]. The $Ca^{2+}$ wave spontaneously originated from a discrete origination site, as reported in *in-vitro* studies, e.g., [19, 33], and propagated away with a constant velocity of 25 μm/s, a value at the center of the physiological *in-vitro* range (8μm/s to 60μm/s) [37-39] which coincides with the reported average wave speed [36].

### b. Spiking Neuron Model

The spiking neurons were simulated as Leaky-Integrate and Fire (LIF) models [48]. The parameters for membrane resistance, *R*, and capacitance, *C*, were adopted from [49, 50] and are given in Table 1.

### c. Astrocyte Model

#### Modeling the Intercellular Ca²⁺ Wave

The intracellular Ca$^{2+}$ level in astrocyte $i$ at time $t$, $c_i(t)$ was modeled by the separable equation:

$$c_i(r_i, t) = a_i * g_i(r_i, t) \tag{1}$$

where $r_i$ is the distance of astrocyte $i$ from the origination site ($r_0 \equiv 0$), $a_i(t)$ and $g_i(t)$ are, respectively, the amplitude and the shape of the *inter*cellular Ca$^{2+}$ wave at astrocyte $i$ and time $t$. Following experimental evidence, we simulated the form of the Ca$^{2+}$ wave, $g_i(r_i, t)$, as the bi-exponential function [51]:

$$g_i(r_i, t) = exp\left(-\frac{r_i - \mu \cdot v \cdot t}{\tau_{decay}}\right) - exp\left(-\frac{r_i - \mu \cdot v \cdot t}{\tau_{rise}}\right) \tag{2}$$

$a_i$ as a decaying exponential:

$$a_i = a_0 * n * exp\left(-\frac{i}{\tau_d}\right), \tag{3}$$

where $\tau_{decay}$, $\tau_{rise}$ were adopted from [29, 51] (see Table 1) and control the rise and fall time of the Ca$^{2+}$ wave; $\tau_d$ controls the magnitude of amplitude fall-off between astrocytes, and *μ* (*μ=1* unless otherwise stated) is a parameter aimed to capture the permeability of the cells through which the wave propagates with speed $v$ --lumping together all the sub-mechanisms of gap-junctional and extracellular communication. The constant *n* is the normalization constant for biexponential distribution, which is a function of the rise and decay parameters (supplementary material). Following modelling studies that showed a relatively small, linear fall-off in amplitude for intercellular Ca$^{2+}$ waves, we made $\tau_d$ comparatively large (Table 1) [38]. We calculated the time it takes for the inter-cellular Ca$^{2+}$ wave peak to propagate between astrocytes $i$ and $i + 1$, $\Delta t_{peak}$, as follows: We first solved for the maxima of $g$:

$$\left.\frac{\partial g_i(r_i,t)}{\partial t}\right|r_j = 0, j = \{i, i+1\}, \tag{4}$$

and then calculated $\Delta t_{peak}$ as:

$$\Delta t_{peak} = t_{peak}^{i+1} - t_{peak}^i = \frac{r_{i+1} - r_i}{\mu v} \tag{5}$$

#### Tripartite Synapse Model

To model the effects of the astrocytes on the synaptic level, we used a modification of the TUM dynamic synapse model [50] known as the Gatekeeper model of the tripartite synapse [22]. Specifically, the dynamics of synaptic vesicle release and recovery in this model are described by:

**Table 1**. Parameter values for the neural-astrocytic network.

| Neuron Parameters [49, 50] | Astrocyte Parameters [29, 30, 36, 52, 53] | Tripartite Synapse Parameters [22, 50] |
|---|---|---|
| $R = .6\ G\Omega$ | $\tau_{decay} = 1.1\ s$ | $\tau_{Ca} = 4\ s$ |
| $C = 100\ pF$ | $\tau_{rise} = 0.5\ s$ | $\tau_{in} = 50\ ms$ |
| $v_{thresh} = 25\ mV$ | $r_{i+1} - r_i = 50\ \mu m$ | $\tau_{rec} = 100\ ms$ |
| $v_{reset} = -65\ mV$ | $\tau_d = 12\ cells$ | $u = 0.1\ (unitless)$ |
|  | $\tau_{decay}^{SIC} = 300ms$ | $\kappa = 0.5\ s$ |
|  | $\tau_{decay}^{SIC} = 50ms$ | $[Ca_{thresh}^{2+}] = 196.69nM$ |
|  | $a_0 = 500nM$ | $A = 1000\ pA$ |

$$\frac{df}{dt} = \left(-\frac{f}{\tau_{Ca}}\right) + (1-f) \cdot \kappa \Theta([Ca^{2+}] - [Ca_{thresh}^{2+}]) \tag{6}$$

$$\frac{dx}{dt} = \frac{z}{\tau_{rec}} - (1-f) \cdot u \cdot x \cdot \delta(t - t_{sp}) - x * \eta(f) \tag{7}$$

$$\frac{dy}{dt} = -\frac{y}{\tau_{in}} + (1-f) \cdot u \cdot x \cdot \delta(t - t_{sp}) + x * \eta(f) \tag{8}$$

$$\frac{dz}{dt} = \frac{y}{\tau_{in}} - \frac{z}{\tau_{rec}} \tag{9}$$

where $f$ is a gating variable that models Ca$^{2+}$-dependent presynaptic inhibition at the tripartite synapse, $\tau_{Ca}$ is the decay constant of $f$, $\kappa$ controls the rise time of $f$, $\Theta(x)$ is the Heaviside function, and $[Ca_{thresh}^{2+}]$ is the concentration of intracellular Ca$^{2+}$, $[Ca^{2+}]$, in a given astrocyte needed to activate $f$. The variables $x, y,$ and $z$ represent the fractions of synaptic resources in a recovered, active, and inactive state, respectively; $\tau_{rec}$ and $\tau_{in}$ are the characteristic times of postsynaptic currents (PSCs) decay, and recovery time from synaptic depression, respectively, and $u$ is the fraction of $x$ released when a spike arrives at the synapse at time $t_{sp}$. The noise term, $\eta(f)$, is a stochastic term introduced into the model to fit the experimentally observed increase of PSCs during a Ca$^{2+}$ wave event and to compensate for the lack of direct NMDAR-mediated effects on the postsynaptic neuron from astrocyte-derived gliotransmission in the Gatekeeper model [22]. The total synaptic current in the gatekeeper model is given by:

$$I_{syn} = A * y(t), \tag{10}$$

where $A$ is a constant multiplicative factor (Table 1), which controls for the network connectivity: When $A = 0$, then $I_{syn} = 0$ and the neurons do not communicate at all; when $A > 0$, pre-synaptic spikes trigger the associated post-synaptic neuron with strength increasing with $A$.

### *Modifications to the Gatekeeper Model*

Since both excitatory and inhibitory tripartite modulations have been implicated in rhythmogenesis [42] ,we directly incorporated excitatory NMDAR-mediated effects into the synapse. To do so, we modified the Gatekeeper model by removing the noise term, $\eta(f)$, in an approach similar to [52], and directly modeled the dynamics of astrocyte-derived NMDAR slow-mediated currents (SICs) into the post-synaptic neuron, as explained below.

The shape, amplitude and time correlations of SICs with respect to astrocyte $Ca^{2+}$ levels were modeled based on experimental data. Specifically, *in-vitro* and *in-vivo* studies reveal that SICs are well fit to bi-exponential distributions, have a rapid rise time (on the order of tens of milliseconds), and a comparatively larger decay time (on the order of hundreds of milliseconds) [27, 53-55]. Furthermore, SICs are correlated with $Ca^{2+}$ wave peaks in both time [56] and amplitude [45, 53]. The amplitude dependence of SICs on astrocyte $Ca^{2+}$ followed the experimentally fit function [45, 53]:

$$I_{astro} = 2.11 \cdot \frac{uA}{cm^2} \cdot ln(w) \cdot \Theta(ln(w)), \tag{11}$$

where

$$w = [Ca^{2+}]/nM - 196.11 \tag{12}$$

and $\Theta$ (x) was the Heaviside function. The total equation for SIC dynamics, activated at every $Ca^{2+}$ peak, was:

$$I_{SIC}(t) = I_{astro}\left([Ca^{2+}] = [Ca^{2+}_{peak}]_i\right) * n * \left(exp\left(-\frac{t}{\tau^{SIC}_{decay}}\right) - exp\left(-\frac{t}{\tau^{SIC}_{rise}}\right)\right), \tag{13}$$

where *n* is a normalization constant for exponential distributions, which assures that the peak of the bi-exponential distribution before being multiplied by $I_{astro}$ is equal to one—a derivation is provided in the supplementary material. In other words, upon reaching a $Ca^{2+}$ peak, an astrocyte releases gliotransmitters that lead to NMDAR mediated SICs, with the SIC amplitude being logarithmically proportional to the $Ca^{2+}$ wave amplitude. Overall, the total current injected into a post-synaptic neuron of a tripartite synapse, $I_{total}$, was the sum of the synaptic current, $I_{syn}$, the astrocyte-derived SIC, $I_{SIC}$, and the basal current $I_{basal}$:

$$I_{total} = I_{syn} + I_{SIC} + I_{basal}. \tag{14}$$

*Modeling Astrocytic Dysfunction*

Astrocytes have long been linked to the pathophysiology of brain diseases [57, 58]. Despite the highly variant experimental results, the intrinsic variability of any brain and the multifaceted complexity of its diseases, a striking observation is that the amplitude as well as the speed of the $Ca^{+2}$ wave are consistently involved in these diseases. While there is no consensus of what constitutes a "physiological" $Ca^{2+}$ wave amplitude *in vivo*, numerous studies support a range for its value to be [250-1000] nM [23, 53, 59]. Therefore, we modeled the effects of astrocyte $Ca^{2+}$ wave amplitude on brain pathophysiology, at least in terms of power differences observed in ISO and SO, by varying the amplitude of the initial $Ca^{2+}$ wave, $a_0$, in the reported range. The Python model code, allowing the full replication of the simulations for Figures 3-6, is available at http://combra.cs.rutgers.edu/astrocyence.

## 4. Results

### a. Network Connectivity and Dynamics

Following the above biological constrains, our neuron-astrocyte network had astrocytes that were radially distributed around a central astrocyte, which initiated a $Ca^{2+}$ wave (Fig. 1a). $Ca^{2+}$ waves can be initiated by an astrocyte either spontaneously, as in our current implementation, or upon sensing the nearby neural activity. Each astrocyte defined a synaptic island, as has been found [31]. The neurons in a particular synaptic island connected in a feedforward manner to the next island (Fig. 1b); Astrocytes were connected via gap junctions (Fig. 1c), forming a fully connected unidirectional network. Neuron-to-neuron connections occurred in the presence of an astrocyte process forming a tripartite synapse (Fig. 1d) [23]. The $Ca^{2+}$ wave decreased in amplitude as it propagated across the 5 astrocytes (Fig. 2 and eqs. 1-3).

### b. *Slow Oscillations Emerging from Astrocyte-Neuron Interactions as a Network Characteristic*

As the $Ca^{2+}$ wave propagated away from the central astrocyte (Fig. 3a), it activated pre-synaptic Adenosine A1 receptors (eqs. 6-9) and post-synaptic NMDAR-mediated SICs (eqs. 12, 13). As the wave propagated along the layers of the network via the synaptic islands, it caused a transient increase in the post-synaptic activity of neurons on each layer (Fig. 3b). An elevation in the neural spiking activity of an island propagated between subsequent islands; this caused a similar transient increase of the spiking activity on all subsequent islands (Fig. 3c), which depended on the strength of neural connectivity between layers. Critically, there were two frequency components, one inside the ISO and the other inside the SO range, that were observed in the firing rate of the network, ranging from the single-neuron to the single-layer to, ultimately, the aggregate spike activity (Fig. 4). The two frequencies emerged from two distinct astrocyte-derived processes, as explained below. In our network, the ISO increased with the frequency of the $Ca^{2+}$ wave origination rate while the SO increased with the speed of $Ca^{2+}$ wave propagation. In this study, we kept the origination rate constant at 0.1 Hz, a value commonly reported in *in-vivo* studies [20]. We also chose a wave speed of 25μm/s, which is in the middle of the physiological range (Table 1). This resulted to an SO of 0.5Hz; Fig. 5a further examines the relationship between the SO frequency and $Ca^{2+}$ wave speed velocity, within the range commonly reported for normal brain tissues, showing the slope of the linear relationship to be equal to the inverse of the inter-astrocyte distance.

### c. *Frequency analysis of aggregate firing activity motivated by pathophysiology*

Pathophysiological brain states exhibit disease-specific changes in oscillations [60] as well as astrocytic dysfunctions. We used our model to explore the potential role of the astrocyte $Ca^{2+}$ wave dynamics in pathophysiological oscillations. In this study, we focused on the effect of relevant model parameters to the oscillations' power, which are commonly imbalanced in various brain diseases.

We first examined how the ISO and SO power depended on two variables that control the synaptic dynamics: $f$, the gating variable, which is related to the activation of the A1 adenosine receptors, and $I_{astro}$, which is the amplitude of the SICs (Fig. 5b-c). When we increased $I_{astro}$, the transient increase in the firing rate became more prominent and this resulted to an SO power increase. The same effect was achieved by increasing the gating strength, $f$, which inhibited transmission of the pre-synaptic spiking activity and, therefore, made the astrocyte-derived SIC signal more prominent (Fig. 5c). The ISO power was mainly related to the gating strength and to a lesser extent to the $I_{astro}$ (Fig. 5b).

We also examined how the ISO and SO power was modulated by network connectivity and the amplitude of the astrocyte $Ca^{2+}$ waves. While the ISO power linearly increased with the network connectivity ($R^2 = 0.99$), the SO power stayed relatively constant (Fig. 6a). The SO and ISO power depended linearly ($R^2 = 0.977, R^2 = 0.95$) on the $Ca^{2+}$ wave amplitude (Fig. 6b). These results are interesting in light of the fact that a multitude of experimental studies show a positive correlation between the amplitude of astrocyte $Ca^{2+}$ waves and slow-/delta-wave power in pathophysiological brain states.

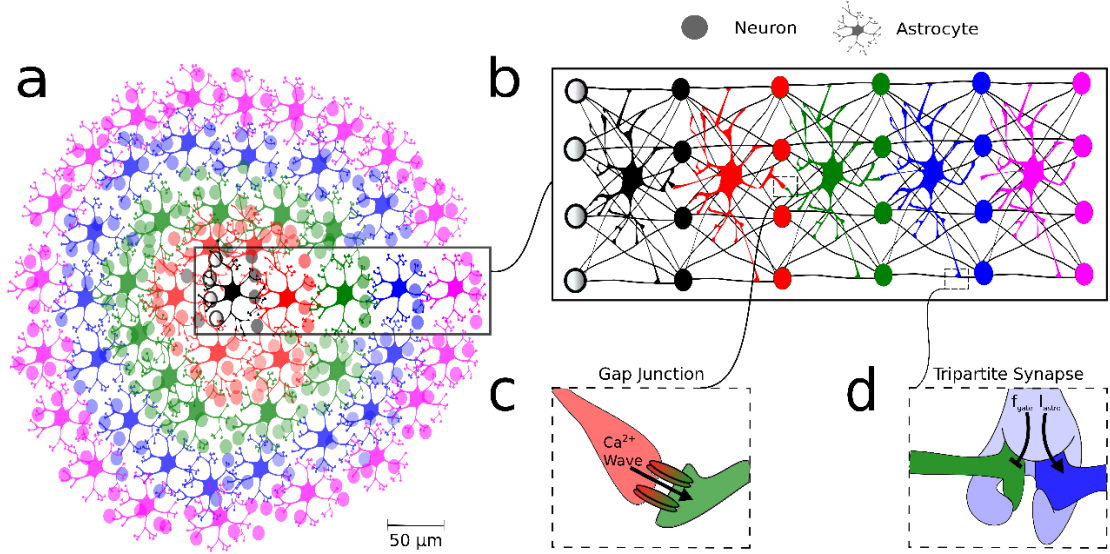

**Figure 1.** *The architecture of the proposed astrocyte-neuron model* a. The characteristic tiling pattern of astrocytes in the brain. The Ca$^{2+}$ wave progenitor astrocyte in denoted in black. 1b. The synaptic island organization of astrocyte-neuron networks. An astrocyte for a given layer entirely envelops up to eight neurons and is associated with up to $10^5$ tripartite synapses c. An astrocyte-astrocyte gap junction; d. A tripartite synapse.

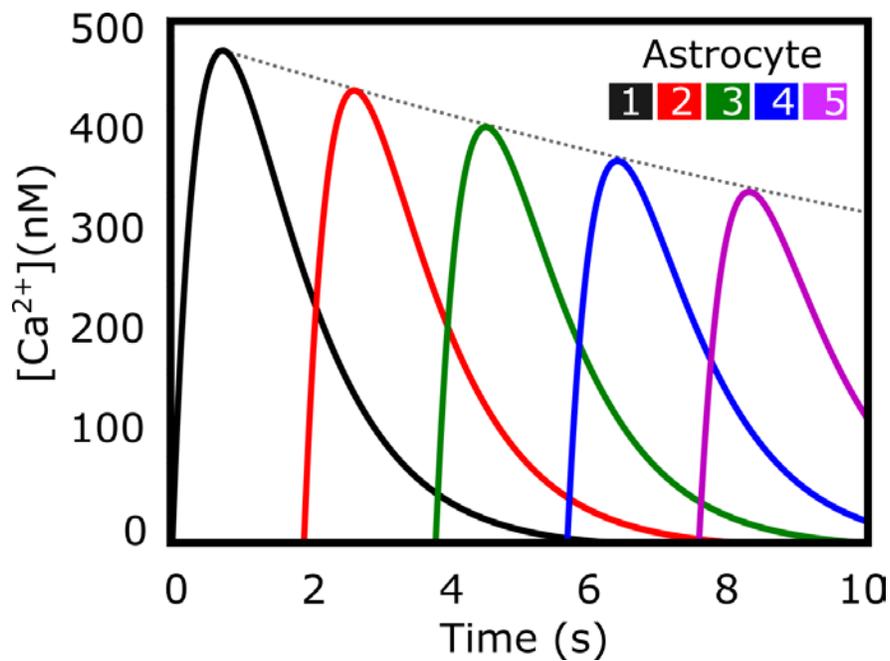

**Figure 2.** Exponential decay of the Ca$^{2+}$ wave amplitude per astrocyte. The peak amplitude, $a_i$, for astrocyte *i* was given by $a_i = a_0 * \exp\left(-\frac{i}{\tau_d}\right)$, where $a_0$ was the initial Ca$^{2+}$ wave amplitude and $\tau_d$ was the fall-off constant. $a_i$ is shown as a dotted line.

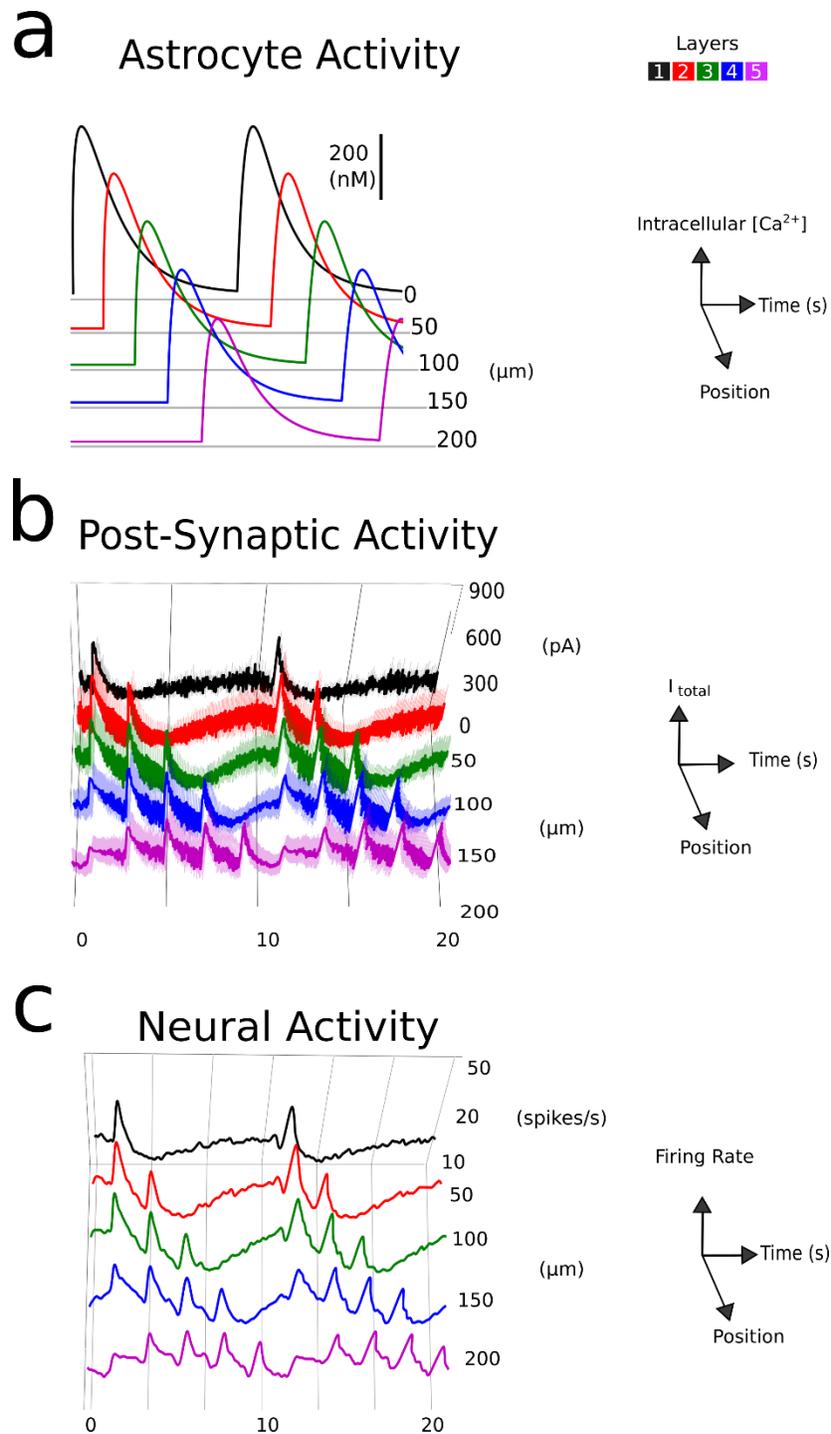

**Figure 3**. Astrocyte $Ca^{2+}$ waves trigger excitatory post-synaptic NMDAR-mediated SICs along the wave path, while simultaneously inhibiting pre-synaptic transmission via Adenosine A1 receptor activation. This results in a rhythmic modulation of neuronal firing patterns, primarily in the slow and infra-slow frequency bands (smoothing method: Hanning Window Length of 1s).

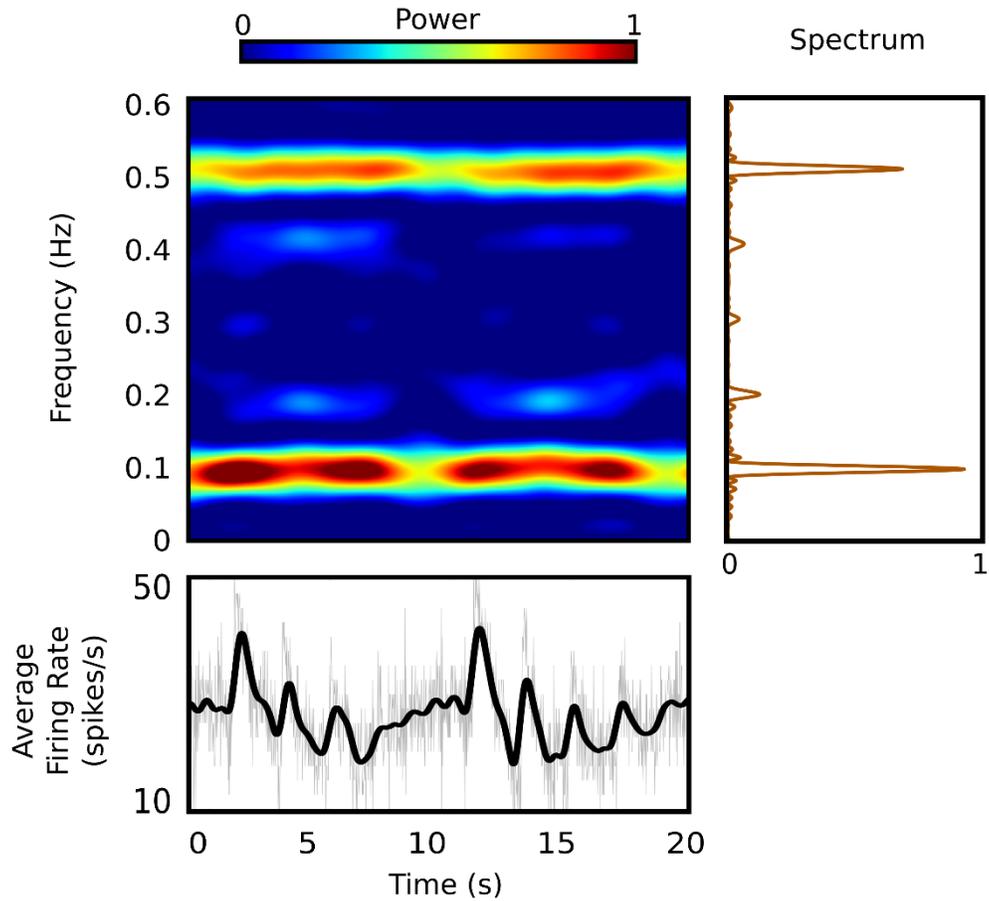

**Figure 4.** Time-series and time-frequency analysis of the aggregate neural network activity indicating two distinct frequencies: the ISO (0.1 Hz) and the SO (0.5 Hz). The ISO corresponds to the frequency of the $Ca^{2+}$ wave initiation for the first astrocyte, while the SO is dependent on the $Ca^{2+}$ wave propagation speed. Bottom subplot: The thick black line is a smooth version of the network firing rate (grey thin line, bin size = 20ms) (Hanning Window Length = 1s); Right subplot: The periodogram of the smoothed firing rate (Length of FFT= $10^6$).

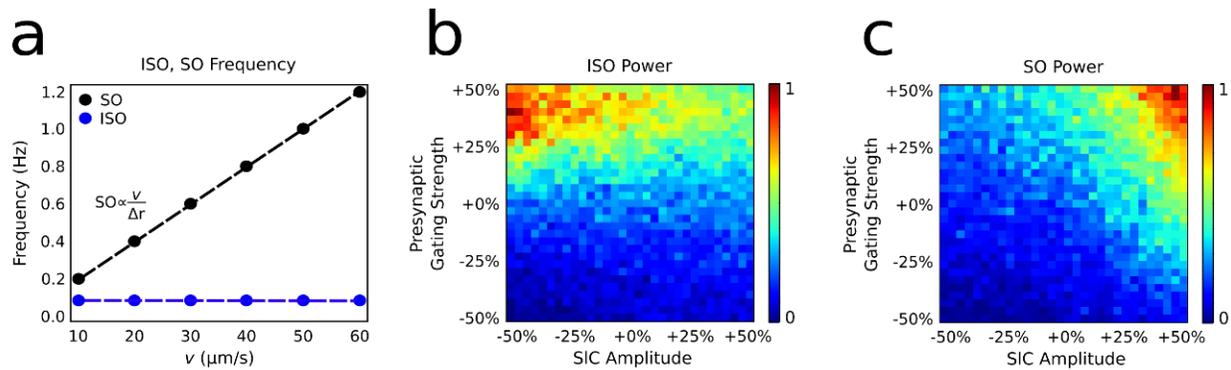

**Figure 5.** a) Linear dependence of peak SO position on velocity; the independence of ISO position on velocity. The slope of the SO line is proportional to the inverse of the inter-astrocyte distance, as expected from theoretical considerations (see eq. 5); the standard error for both regressions was < $10^{-6}$ Hz. b) The ISO power as a function of presynaptic gating strength and SIC amplitude, showing the dominance of the SO over frequency space at high SIC amplitudes. c) The same plot for SO. As the presynaptic signal is gated, the astrocyte-derived SIC is expressed more clearly.

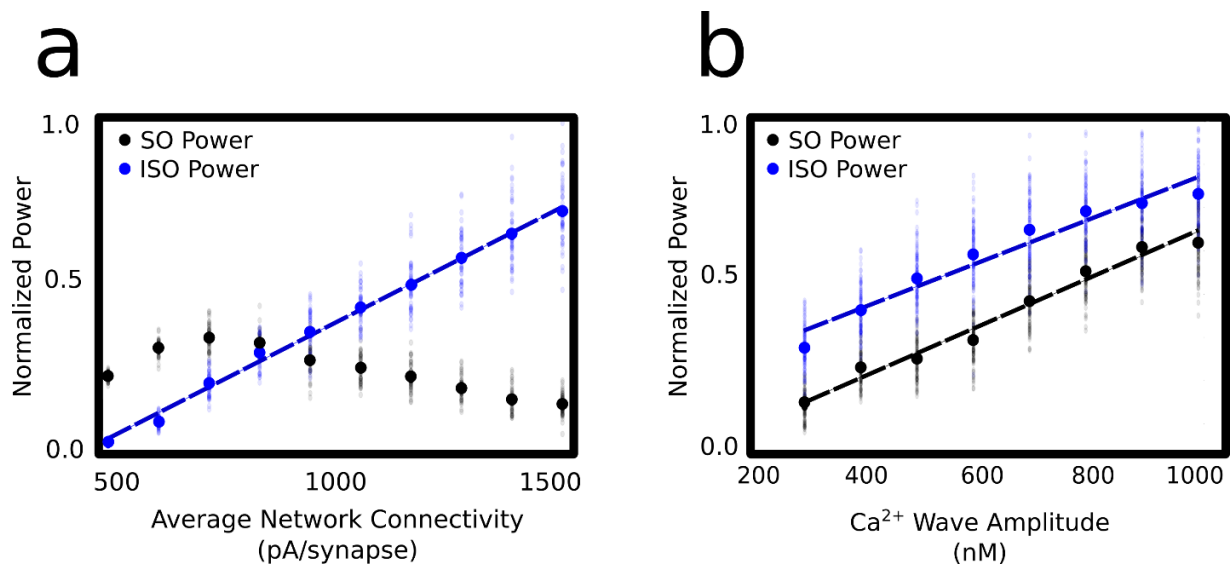

**Figure 6.** The dependence of the ISO and the SO on a) network connectivity and b) $Ca^{2+}$ wave amplitude, motivated by brain pathophysiology.

## 5. Discussion

Our study proposes a mechanistic explanation for the long-suspected causal role of astrocytes on oscillations in the infra-slow and slow frequency bands. The main contribution of this work is that ISO and SO, two prominent and wide-spread slow oscillations modulating faster oscillations, arise as a property of neural-astrocytic network interactions. Our neurophysiologically plausible model suggests that these interactions are influenced by factors such as astrocyte connectomics, $Ca^{2+}$ wave dynamics, and synaptic efficiency. As a validation method for our model, we show that it can consistently support a linkage of the widely reported oscillatory imbalances in brain disorders to known astrocytic dysfunctions in these diseases.

Questions concerning the origin and the mechanisms of slow oscillations arose even from the very first observation of their presence in the cortex [61]. In cortical areas, slow oscillations appear to be generated locally, i.e. "in cortico-cortical networks, since they survive thalamectomy, but not the disruption of cortico-cortical connections" and are further found to modulate "delta, theta, spindle, alpha, beta, gamma and ripple oscillations" [62]. In healthy humans, each individual slow wave cycle has a distinct origin and propagates uniquely across brain areas [63]; Similar patterns were also found in animal models [64, 65]. These findings corroborate that slow oscillations emerge as a network mechanism radiating across large brain areas and starting from distinct sources, a structural and functional pattern that is consistent with that found in astrocytes.

Indirect evidence on how astrocytes participate in the brain's computations suggests that they organize neurons into functional, synchronous clusters, and integrate information across large brain regions [43]. Alongside the structural similarity, the similarity in frequency profiles of astrocytic $Ca^{2+}$ activity and neural spiking activity during ISOs is striking. Lőrincz et al. have reported thalamic astrocytes exhibiting *in situ* spontaneous slowly propagating intracellular $Ca^{2+}$ oscillations in an almost identical range of frequencies (0.003–0.1 Hz) to the neuronal ISO, given rise by a slow 'wave-like' co-activation of different neurons and a sometimes biphasic nature of the long-lasting hyperpolarizing potentials [15]. The extent of the above similarities spreading across the structural and functional domains has rarely been overlooked as a coincidence.

In fact, the role of astrocytes on brain function has long been speculated [66]. More than half a century ago, Robert Galambos recognized the potential of "an alternate theory in which a glia-neural complex of cells forming the functional unit" of brain computation [67]. In recent years, both astrocytes and slow oscillations have, in separate studies, been implicated in cognitive performance: Stimulating slow wave oscillations via transcranial stimulation is known to improve performance in memory tasks [68]; it is also known to increase astrocyte $Ca^{2+}$ activity [69]. Interestingly, when human glial progenitor cells were engrafted into the forebrains of neonatal rodents, the maturated recipient rodent brain "exhibited large numbers and high proportions of both human glial progenitors and astrocytes" and, consequently, outperformed their normal peers significantly on memory and learning tasks [70]. A potential explanation for the observed enhancement of cognitive abilities is that astrocyte might evoke a phase-of-firing modulation of neural activity by astrocytes [71]. Our model aligns with this hypothesis as astrocytes can increase the firing of neurons at a certain phase within a period (Fig. 3) and also increase the SO and ISO power by increasing their $Ca^{2+}$ wave amplitude (Fig. 6b): This transient increase of firing rate could facilitate the firing of the postsynaptic neurons for a given input by decreasing the entropy of the signal passing through the synapses at these moments.

Several experimental studies have targeted astrocytes as a possible source of slow-wave neural oscillations [15, 42, 43]. In agreement with our modeling approach (eq. 11-14; Fig. 5b-c), these studies have emphasized the contribution of astrocytes on simultaneous presynaptic inhibition and postsynaptic excitation and further suggested that their combination might generate the slow wave oscillations. Astrocytes have also been associated with blood-oxygen-level-dependent (BOLD) signals, which govern the ISO and SO frequency range, both in resting-state [3, 72, 73], and stimulus-evoked functional magnetic resonance imaging (fMRI) studies [74]. Interestingly, a change in gliotransmission is found to significantly alter the power of slow-wave oscillations [42] and, relatedly, modulate cortical UP and DOWN states [7]. Interestingly enough, our model suggests that the astrocytic network can modulate the firing rate at the SO range by simply varying the inter-cellular $Ca^{2+}$ wave propagation speed (Fig. 5a). This proposes a potential mechanism where different astrocytic networks can change their own $Ca^{2+}$ wave propagation speed to allow neural encoding of information through amplitude, frequency or amplitude-frequency modulation, a hypothesis that has been proposed by previous computational studies [16].

The type of generation mechanism that we propose here seems to be also present in glia-like support cells, which behave similarly to astrocytes. During early development, glia-like inner support cells within Kolliker's organ spontaneously generate $Ca^{2+}$ waves which propagate radially outward at a speed comparable to that of astrocytes [75]. They are also organized into clusters roughly 60μm apart [76], close to the distance between synaptic islands (50μm). Notably, neurons within the developing auditory system were found to receive slow inward currents with a frequency of roughly 0.2Hz; these SIC's were coincident with inner support cell $Ca^{2+}$ wave events 85% of the time [75]. The reported $Ca^{2+}$ waves speeds were between 5 and 15 μm/s. Based on our model (eq. 5; Fig. 5a), for an inter-astrocyte distance of 60μm, a frequency of 0.2 Hz would be generated by a $Ca^{2+}$ wave propagation speed of 12 μm/s, which is within the experimental reported range of values.

### *Implications to brain pathophysiology and its treatment*

Alongside the mounting evidence that neuron-astrocyte interaction plays an important role in brain function, many studies are also suggesting that a dysfunction in their interactions may accompany neurological disorders. On one hand, the most commonly reported astrocytic dysfunction is $Ca^{2+}$ wave amplitude changes; on the other hand, the most commonly examined slow-oscillation imbalances are power changes. Our model suggests that an increase of the $Ca^{2+}$ wave amplitude results to an SO and ISO power increase (Fig. 6b); it further suggests that increasing the connectivity among neurons would cause the ISO, but not the SO, power to increase (Fig. 6a). ISO and SO power changes can also be achieved through intrinsic astrocytic or synaptic mechanisms which would affect the SIC amplitude or the presynaptic gating strength (Fig. 5b-c). It is interesting to see how well our modeling results align with experimental studies that have examined, separately, how brain disorders affect the astrocytic $Ca^{2+}$ wave changes and the oscillatory imbalances.

Evidence for astrocyte involvement in major depression disorder (MDD) suggests that the amplitude of astrocyte $Ca^{2+}$ waves is altered as a result of the illness. Specifically, depressive rats exhibit decreased levels of ATP, which is known to trigger and enhance astrocyte $Ca^{2+}$ waves [77]; in line with our modeling results, depressive patients exhibit lower power in delta/slow wave activity than the normal population [78, 79]. A striking observation is that trans-cranial stimulation, when used to alleviate the symptoms of MDD, is reported to both increase delta/slow oscillation activity and increase astrocyte $Ca^{2+}$ activity in separate studies [69, 80].

The causal role of astrocytes in the onset of AD and epilepsy has long been speculated and confirmed [81-85]. In AD, the cognitive loss is linked to the dysfunction of synaptic plasticity which is known to be maintained, at least partially, by the astrocytes. A pathological hallmark of AD is the increased levels of β-Amyloid (Aβ) [86-88] which is associated with a decrease of the astrocytic capacity to internalize Aβ [89, 90]. The effect of blocking astrocytic purinergic receptors has long been known to cause a reduction in $Ca^{2+}$ wave velocity and amplitude, resulting to a significant decrease in the wave spread [91]. Studies have shown that cultured astrocytes respond to elevation in environmental Aβ by increasing the speed and amplitude of their calcium waves: the former by up to 280% and the latter by up to 207% [92]. In alignment with our modeling results, a multitude of studies report an increase of slow oscillations in AD; for a review see [93]. Interestingly, some of these studies include results for the ISO and the SO (low delta) band, e.g., [94-96]. Similar to AD, astrocytes in epileptic brain tissue also exhibit velocities and amplitudes above the normal range. For instance, studies in which seizures were induced by the application of 4-Aminopyridine (4-AP), a known convulsant, found a significant increase in astrocyte activity [84]. This brain disease, too, is typically accompanied by an increase of the SO power, at least for temporal lobe epilepsy [97] and the same side of the seizure source [98].

Two other astrocyte-implicated brain diseases are ischemic stroke and PD. In the case of the acute ischemic stroke, a significant [100-500]% increase in the astrocyte $Ca^{2+}$ activity has been reported in the nearby brain tissue with the peak value taken 80 minutes after photothrombosis [99]. The same disease has been related to a [63-120]% increase of the delta/SO power, at least for the most severe cases [100]. Involving deep brain structures, PD cannot faithfully be linked to cortical oscillations. Nevertheless, initial epidemiological studies on the relationship of the disease and caffeine consumption, further supported by experimental studies on adenosine neurotransmission, have revealed astrocytes as a key mechanism that can affect the onset or progression of PD [101]. In an animal model of PD, astrocytes located in the Globus Pallidus externus have been found to exhibit increased $Ca^{2+}$ waves in both frequency and amplitude [102]. Based on our model, an increase of the $Ca^{2+}$ wave results to an SO and ISO power increase (Fig. 6b). Awake EEG studies in patients with PD have shown an increase in slow delta activity across the cortex, with the oscillation increasing with cognitive impairment [103-105]. Conspicuously, adenosine receptors on astrocytes have recently been proposed as potential targets for the treatment of PD [106]. To speculate further, since PD is a network disease, as it affects a multitude of closely connected nuclei, it is not hard to imagine a neuroprotective approach that takes into account astrocytes as a functional unit integrated into the neural network, and not just as a cell that mediates neuronal survival and therefore has a limited number of possible interventions [107].

Overall, our modeling results are in agreement with the experimental studies, where slow-wave oscillations are altered in astrocyte-implicated brain disorders. Whether our model can be used to shed light on the astrocytes' role on brain diseases remains to be seen.

## 6. Conclusion

Slow-wave oscillations, whose precise origin and function are not known, are considered a fundamental mode of brain activity which is generalizable across cortical areas and species and typically appears when the brain enters a resting-state, such as sleep or coma. Astrocytes have long been known to be actively involved in neural circuits necessary for key brain functions, including sleep, memory, and sensory processing. In this paper, we presented a biologically plausible computational model of a simple astrocytic-neural network to propose a possible generation mechanism for ISO and SO. The model qualitatively predicts the slow oscillation imbalances seen in astrocyte-implicated brain disorders. This and other efforts aiming to reveal the computational role of the astrocytes in brain function and dysfunction have the potential to shape a new research field, which one could call "computational astrocyence."

## 7. Appendix

### *Derivation of Biexponential Normalization Constant*

Let $b(x)$ be a biexponential function with time constants $a$ and $b$ and normalization constant $n$:

$$b(x) = n * (exp\left(-\frac{x}{b}\right) - exp\left(-\frac{x}{a}\right))$$

We wish to solve for $n$ such that $b(x_{max}) = 1$. To determine $x_{max}$, we take the partial derivative of $b(x)$ with respect to x and set this equal to zero:

$$\frac{\partial b}{\partial x} = n * \left(-\left(\frac{1}{b}\right) * \exp\left(-\frac{x_{max}}{b}\right) + \left(\frac{1}{a}\right) * \exp\left(-\frac{x_{max}}{a}\right)\right) = 0$$

$$x_{max} = (\ln\left(\frac{b}{a}\right) * a * b) * (b - a)^{-1}$$

Plugging $x_{max}$ back into $b(x)$ gives us the maximum amplitude of the function:

$$b(x_{max}) = n * (\left(\frac{b}{a}\right)^{\frac{a}{a-b}} - \left(\frac{b}{a}\right)^{\frac{b}{a-b}})$$

Therefore, the value of $n$ which assures that $b(x_{max}) = 1$ is given by:

$$n = \left(\left(\frac{b}{a}\right)^{\frac{a}{a-b}} - \left(\frac{b}{a}\right)^{\frac{b}{a-b}}\right)^{-1}$$

In the case of $[a = 50, b = 300]$ we get $n = \frac{6 * 6^{\frac{1}{5}}}{5}$.

## 8. Acknowledgments

We would like to thank Kevin T. Feigelis for his insightful suggestions and modeling work during the early stages of this project.

## 9. References

1. Sejnowski TJP, O. Network Oscillations: Emerging Computational Principles. Journal of Neuroscience Journal of Neuroscience 2006;26:1673–6
2. Steriade M. Grouping of brain rhythms in corticothalamic systems. Neuroscience. 2006;137(4):1087-106.
3. Hiltunen T, Kantola J, Elseoud AA, Lepola P, Suominen K, Starck T, et al. Infra-slow EEG fluctuations are correlated with resting-state network dynamics in fMRI. The Journal of Neuroscience. 2014;34(2):356-62.

4. Buckner RL, Andrews-Hanna JR, Schacter DL. The brain's default network. Annals of the New York Academy of Sciences. 2008;1124(1):1-38.
5. Buzsáki G, Chrobak JJ. Temporal structure in spatially organized neuronal ensembles: a role for interneuronal networks. Current opinion in neurobiology. 1995;5(4):504-10.
6. Nir Y, Staba RJ, Andrillon T, Vyazovskiy VV, Cirelli C, Fried I, et al. Regional slow waves and spindles in human sleep. Neuron. 2011;70(1):153-69.
7. Poskanzer KE, Yuste R. Astrocytic regulation of cortical UP states. Proceedings of the National Academy of Sciences. 2011;108(45):18453-8.
8. McGinty D, Szymusiak R. Keeping cool: a hypothesis about the mechanisms and functions of slow-wave sleep. Trends in neurosciences. 1990;13(12):480-7.
9. Jones S. Slow-wave sleep and the risk of type 2 diabetes in humans Tasali E, Leproult R, Ehrmann DA, et al (Univ of Chicago, IL) Proc Natl Acad Sci USA 105: 1044-1049, 2008. Year Book of Pulmonary Disease. 2009;2009:277-8.
10. Buzsaki G. Rhythms of the Brain: Oxford University Press; 2006.
11. Helkala E-L, Laulumaa V, Soikkeli R, Partanen J, Soininen H, Riekkinen PJ. Slow-wave activity in the spectral analysis of the electroencephalogram is associated with cortical dysfunctions in patients with Alzheimer's disease. Behavioral neuroscience. 1991;105(3):409.
12. Göder R, Boigs M, Braun S, Friege L, Fritzer G, Aldenhoff JB, et al. Impairment of visuospatial memory is associated with decreased slow wave sleep in schizophrenia. Journal of psychiatric research. 2004;38(6):591-9.
13. Herculano-Houzel S. The human brain in numbers: a linearly scaled-up primate brain. Frontiers in human neuroscience. 2009;3:31.
14. Verkhratsky A, Butt A. Neuroglia: definition, classification, evolution, numbers, development. glial physiology and pathophysiology. 2013:73-104.
15. Lőrincz ML, Geall F, Bao Y, Crunelli V, Hughes SW. ATP-dependent infra-slow (< 0.1 Hz) oscillations in thalamic networks. PLoS One. 2009;4(2):e4447.
16. Goldberg M, De Pittà M, Volman V, Berry H, Ben-Jacob E. Nonlinear gap junctions enable long-distance propagation of pulsating calcium waves in astrocyte networks. PLoS Comput Biol. 2010;6(8):e1000909.
17. Khakh BS, McCarthy KD. Astrocyte calcium signaling: from observations to functions and the challenges therein. Cold Spring Harb Perspect Biol. 2015;7:a020404.
18. Kurth-Nelson ZL, Mishra A, Newman EA. Spontaneous glial calcium waves in the retina develop over early adulthood. The Journal of Neuroscience. 2009;29(36):11339-46.
19. Parri HR, Gould TM, Crunelli V. Spontaneous astrocytic Ca2+ oscillations in situ drive NMDAR-mediated neuronal excitation. Nature neuroscience. 2001;4(8):803-12.
20. Hirase H, Qian L, Barthó P, Buzsáki G. Calcium dynamics of cortical astrocytic networks in vivo. PLoS Biol. 2004;2(4):e96.
21. Bazargani N, Attwell D. Astrocyte calcium signaling: the third wave. Nature neuroscience. 2016;19(2):182-9.
22. Volman V, Ben-Jacob E, Levine H. The astrocyte as a gatekeeper of synaptic information transfer. Neural computation. 2007;19(2):303-26.
23. Araque A, Parpura V, Sanzgiri RP, Haydon PG. Tripartite synapses: glia, the unacknowledged partner. Trends in neurosciences. 1999;22(5):208-15.
24. Florian C, Vecsey CG, Halassa MM, Haydon PG, Abel T. Astrocyte-derived adenosine and A1 receptor activity contribute to sleep loss-induced deficits in hippocampal synaptic plasticity and memory in mice. The Journal of neuroscience. 2011;31(19):6956-62.

25. Lee HS, Ghetti A, Pinto-Duarte A, Wang X, Dziewczapolski G, Galimi F, et al. Astrocytes contribute to gamma oscillations and recognition memory. Proceedings of the National Academy of Sciences. 2014;111(32):E3343-E52.
26. Perea G, Navarrete M, Araque A. Tripartite synapses: astrocytes process and control synaptic information. Trends in neurosciences. 2009;32(8):421-31.
27. Araque A, Parpura V, Sanzgiri RP, Haydon PG. Glutamate-dependent astrocyte modulation of synaptic transmission between cultured hippocampal neurons. European Journal of Neuroscience. 1998;10(6):2129-42.
28. Clarke LE, Barres BA. Emerging roles of astrocytes in neural circuit development. Nature Reviews Neuroscience. 2013;14(5):311-21.
29. De Pitta M, Volman V, Berry H, Parpura V, Volterra A, Ben-Jacob E. Computational quest for understanding the role of astrocyte signaling in synaptic transmission and plasticity. Frontiers in computational neuroscience. 2012;6.
30. Sul J-Y, Orosz G, Givens RS, Haydon PG. Astrocytic connectivity in the hippocampus. Neuron glia biology. 2004;1(01):3-11.
31. Halassa MM, Fellin T, Takano H, Dong J-H, Haydon PG. Synaptic islands defined by the territory of a single astrocyte. The Journal of neuroscience. 2007;27(24):6473-7.
32. Cornell-Bell A, Finkbeiner S. Ca2+ waves in astrocytes. Cell calcium. 1991;12(2-3):185-204.
33. Hoogland TM, Kuhn B, Göbel W, Huang W, Nakai J, Helmchen F, et al. Radially expanding transglial calcium waves in the intact cerebellum. Proceedings of the National Academy of Sciences. 2009;106(9):3496-501.
34. Newman EA. Propagation of intercellular calcium waves in retinal astrocytes and Müller cells. The Journal of Neuroscience. 2001;21(7):2215-23.
35. Newman EA, Zahs KR. Calcium waves in retinal glial cells. Science. 1997;275(5301):844-7.
36. Scemes E, Giaume C. Astrocyte calcium waves: what they are and what they do. Glia. 2006;54(7):716-25.
37. Nimmerjahn A, Bergles DE. Large-scale recording of astrocyte activity. Current opinion in neurobiology. 2015;32:95-106.
38. Höfer T, Venance L, Giaume C. Control and plasticity of intercellular calcium waves in astrocytes: a modeling approach. The Journal of neuroscience. 2002;22(12):4850-9.
39. Kuga N, Sasaki T, Takahara Y, Matsuki N, Ikegaya Y. Large-scale calcium waves traveling through astrocytic networks in vivo. The Journal of Neuroscience. 2011;31(7):2607-14.
40. Nett WJ, Oloff SH, McCarthy KD. Hippocampal astrocytes in situ exhibit calcium oscillations that occur independent of neuronal activity. Journal of neurophysiology. 2002;87(1):528-37.
41. Verderio C, Bacci A, Coco S, Pravettoni E, Fumagalli G, Matteoli M. Astrocytes are required for the oscillatory activity in cultured hippocampal neurons. European Journal of Neuroscience. 1999;11(8):2793-800.
42. Fellin T, Halassa MM, Terunuma M, Succol F, Takano H, Frank M, et al. Endogenous nonneuronal modulators of synaptic transmission control cortical slow oscillations in vivo. Proceedings of the National Academy of Sciences. 2009;106(35):15037-42.
43. Poskanzer KE, Yuste R. Astrocytes regulate cortical state switching in vivo. Proceedings of the National Academy of Sciences. 2016;113(19):E2675-E84.
44. Nadjar A, Blutstein T, Aubert A, Laye S, Haydon PG. Astrocyte-derived adenosine modulates increased sleep pressure during inflammatory response. Glia. 2013;61(5):724-31.
45. Nadkarni S, Jung P. Dressed neurons: modeling neural–glial interactions. Physical biology. 2004;1(1):35.
46. Fourcaud N, Brunel N. Dynamics of the firing probability of noisy integrate-and-fire neurons. Neural computation. 2002;14(9):2057-110.

47. Bennett MR, Farnell L, Gibson WG, Lagopoulos J. Cortical Network Models of Firing Rates in the Resting and Active States Predict BOLD Responses. PloS one. 2015;10(12):e0144796.
48. Lapicque L. Recherches quantitatives sur l'excitation électrique des nerfs traitée comme une polarisation. J Physiol Pathol Gen. 1907;9(1):620-35.
49. Tal D, Schwartz EL. Computing with the leaky integrate-and-fire neuron: logarithmic computation and multiplication. Neural Computation. 1997;9(2):305-18.
50. Tsodyks M, Pawelzik K, Markram H. Neural networks with dynamic synapses. Neural computation. 1998;10(4):821-35.
51. Winship IR, Plaa N, Murphy TH. Rapid astrocyte calcium signals correlate with neuronal activity and onset of the hemodynamic response in vivo. The Journal of neuroscience. 2007;27(23):6268-72.
52. Wade JJ, McDaid LJ, Harkin J, Crunelli V, Kelso JS. Bidirectional coupling between astrocytes and neurons mediates learning and dynamic coordination in the brain: a multiple modeling approach. PloS one. 2011;6(12):e29445.
53. Parpura V, Haydon PG. Physiological astrocytic calcium levels stimulate glutamate release to modulate adjacent neurons. Proceedings of the National Academy of Sciences. 2000;97(15):8629-34.
54. Fellin T, Pascual O, Gobbo S, Pozzan T, Haydon PG, Carmignoto G. Neuronal synchrony mediated by astrocytic glutamate through activation of extrasynaptic NMDA receptors. Neuron. 2004;43(5):729-43.
55. Haydon PG, Carmignoto G. Astrocyte control of synaptic transmission and neurovascular coupling. Physiological reviews. 2006;86(3):1009-31.
56. Pasti L, Zonta M, Pozzan T, Vicini S, Carmignoto G. Cytosolic calcium oscillations in astrocytes may regulate exocytotic release of glutamate. The Journal of Neuroscience. 2001;21(2):477-84.
57. Eddleston M, Mucke L. Molecular profile of reactive astrocytes—implications for their role in neurologic disease. Neuroscience. 1993;54(1):15-36.
58. Barres BA. The mystery and magic of glia: a perspective on their roles in health and disease. Neuron. 2008;60(3):430-40.
59. Fiacco TA, McCarthy KD. Astrocyte calcium elevations: properties, propagation, and effects on brain signaling. Glia. 2006;54(7):676-90.
60. Schnitzler A, Gross J. Normal and pathological oscillatory communication in the brain. Nature reviews neuroscience. 2005;6(4):285-96.
61. Steriade M, Nunez A, Amzica F. A novel slow (< 1 Hz) oscillation of neocortical neurons in vivo: depolarizing and hyperpolarizing components. Journal of neuroscience. 1993;13(8):3252-65.
62. Csercsa R, Dombovári B, Fabó D, Wittner L, Erőss L, Entz L, et al. Laminar analysis of slow wave activity in humans. Brain. 2010:awq169.
63. Murphy K, Birn RM, Handwerker DA, Jones TB, Bandettini PA. The impact of global signal regression on resting state correlations: are anti-correlated networks introduced? Neuroimage. 2009;44(3):893-905.
64. Mohajerani MH, McVea DA, Fingas M, Murphy TH. Mirrored bilateral slow-wave cortical activity within local circuits revealed by fast bihemispheric voltage-sensitive dye imaging in anesthetized and awake mice. Journal of Neuroscience. 2010;30(10):3745-51.
65. Ferezou I, Haiss F, Gentet LJ, Aronoff R, Weber B, Petersen CC. Spatiotemporal dynamics of cortical sensorimotor integration in behaving mice. Neuron. 2007;56(5):907-23.
66. Desmedt JE, La Grutta G. The effect of selective inhibition of pseudocholinesterase on the spontaneous and evoked activity of the cat's cerebral cortex. The Journal of physiology. 1957;136(1):20.
67. Galambos R. A glia-neural theory of brain function. Proceedings of the National Academy of Sciences. 1961;47(1):129-36.

68. Marshall L, Helgadóttir H, Mölle M, Born J. Boosting slow oscillations during sleep potentiates memory. Nature. 2006;444(7119):610-3.
69. Monai H, Ohkura M, Tanaka M, Oe Y, Konno A, Hirai H, et al. Calcium imaging reveals glial involvement in transcranial direct current stimulation-induced plasticity in mouse brain. Nature communications. 2016;7.
70. Han X, Chen M, Wang F, Windrem M, Wang S, Shanz S, et al. Forebrain engraftment by human glial progenitor cells enhances synaptic plasticity and learning in adult mice. Cell stem cell. 2013;12(3):342-53.
71. Zucker SW. Local field potentials and border ownership: a conjecture about computation in visual cortex. Journal of Physiology-Paris. 2012;106(5):297-315.
72. Figley CR, Stroman PW. The role (s) of astrocytes and astrocyte activity in neurometabolism, neurovascular coupling, and the production of functional neuroimaging signals. European Journal of Neuroscience. 2011;33(4):577-88.
73. Palva JM, Palva S. Infra-slow fluctuations in electrophysiological recordings, blood-oxygenation-level-dependent signals, and psychophysical time series. Neuroimage. 2012;62(4):2201-11.
74. Schulz K, Sydekum E, Krueppel R, Engelbrecht CJ, Schlegel F, Schröter A, et al. Simultaneous BOLD fMRI and fiber-optic calcium recording in rat neocortex. Nature methods. 2012;9(6):597-602.
75. Tritsch NX, Yi E, Gale JE, Glowatzki E, Bergles DE. The origin of spontaneous activity in the developing auditory system. Nature. 2007;450(7166):50-5.
76. Dayaratne M, Vlajkovic SM, Lipski J, Thorne PR. Kölliker's organ and the development of spontaneous activity in the auditory system: Implications for hearing dysfunction. BioMed research international. 2014;2014.
77. Cao X, Li L-P, Wang Q, Wu Q, Hu H-H, Zhang M, et al. Astrocyte-derived ATP modulates depressive-like behaviors. Nature medicine. 2013;19(6):773-7.
78. Meerwijk EL, Ford JM, Weiss SJ. Resting-state EEG delta power is associated with psychological pain in adults with a history of depression. Biological psychology. 2015;105:106-14.
79. Brenner RP, Ulrich RF, Spiker DG, Sclabassi RJ, Reynolds CF, Marin RS, et al. Computerized EEG spectral analysis in elderly normal, demented and depressed subjects. Electroencephalography and clinical neurophysiology. 1986;64(6):483-92.
80. Massimini M, Ferrarelli F, Esser SK, Riedner BA, Huber R, Murphy M, et al. Triggering sleep slow waves by transcranial magnetic stimulation. Proceedings of the National Academy of Sciences. 2007;104(20):8496-501.
81. Seifert G, Carmignoto G, Steinhäuser C. Astrocyte dysfunction in epilepsy. Brain research reviews. 2010;63(1):212-21.
82. de Lanerolle NC, Lee T-S, Spencer DD. Astrocytes and epilepsy. Neurotherapeutics. 2010;7(4):424-38.
83. Hinterkeuser S, SchroÈder W, Hager G, Seifert G, BluÈmcke I, Elger CE, et al. Astrocytes in the hippocampus of patients with temporal lobe epilepsy display changes in potassium conductances. European Journal of Neuroscience. 2000;12(6):2087-96.
84. Tian G-F, Azmi H, Takano T, Xu Q, Peng W, Lin J, et al. An astrocytic basis of epilepsy. Nature medicine. 2005;11(9):973-81.
85. Verkhratsky A, Olabarria M, Noristani HN, Yeh C-Y, Rodriguez JJ. Astrocytes in Alzheimer's disease. Neurotherapeutics. 2010;7(4):399-412.
86. Scheuner D, Eckman C, Jensen M, Song X, Citron M, Suzuki N, et al. Secreted amyloid beta-protein similar to that in the senile plaques of Alzheimer's disease is increased in vivo by the presenilin 1 and 2 and APP mutations linked to familial Alzheimer's disease. Nature medicine. 1996;2(8):864-70.
87. Duff K, Eckman C, Zehr C, Yu X, Prada C-M, Perez-Tur J, et al. Increased amyloid-β42 (43) in brains of mice expressing mutant presenilin 1. Nature. 1996;383(6602):710-3.

88. Schmechel D, Saunders A, Strittmatter W, Crain BJ, Hulette C, Joo S, et al. Increased amyloid beta-peptide deposition in cerebral cortex as a consequence of apolipoprotein E genotype in late-onset Alzheimer disease. Proceedings of the National Academy of Sciences. 1993;90(20):9649-53.
89. Zhao J, O'Connor T, Vassar R. The contribution of activated astrocytes to Aβ production: implications for Alzheimer's disease pathogenesis. Journal of neuroinflammation. 2011;8(1):150.
90. Pihlaja R, Koistinaho J, Kauppinen R, Sandholm J, Tanila H, Koistinaho M. Multiple cellular and molecular mechanisms are involved in human Aβ clearance by transplanted adult astrocytes. Glia. 2011;59(11):1643-57.
91. Scemes E, Dermietzel R, Spray DC. Calcium waves between astrocytes from Cx43 knockout mice. Glia. 1998;24(1):65.
92. Haughey NJ, Mattson MP. Alzheimer's amyloid β-peptide enhances ATP/gap junction-mediated calcium-wave propagation in astrocytes. Neuromolecular medicine. 2003;3(3):173-80.
93. Jeong J. EEG dynamics in patients with Alzheimer's disease. Clinical neurophysiology. 2004;115(7):1490-505.
94. Hier DB, Mangone CA, Ganellen R, Warach JD, Van Egeren R, Perlik SJ, et al. Quantitative measurement of delta activity in Alzheimer's disease. Clinical EEG and Neuroscience. 1991;22(3):178-82.
95. Fonseca LC, Tedrus GMAS, Prandi LR, Andrade A. Quantitative electroencephalography power and coherence measurements in the diagnosis of mild and moderate Alzheimer's disease. Arquivos de neuro-psiquiatria. 2011;69(2B):297-303.
96. Neto E, Allen EA, Aurlien H, Nordby H, Eichele T. EEG spectral features discriminate between Alzheimer's and vascular dementia. Frontiers in neurology. 2015;6.
97. Bernasconi A, Cendes F, Lee J, Reutens DC, Gotman J. EEG background delta activity in temporal lobe epilepsy: correlation with volumetric and spectroscopic imaging. Epilepsia. 1999;40(11):1580-6.
98. Englot DJ, Yang L, Hamid H, Danielson N, Bai X, Marfeo A, et al. Impaired consciousness in temporal lobe seizures: role of cortical slow activity. Brain. 2010;133(12):3764-77.
99. Ding S. Ca2+ signaling in astrocytes and its role in ischemic stroke. Glutamate and ATP at the Interface of Metabolism and Signaling in the Brain: Springer; 2014. p. 189-211.
100. Finnigan SP, Rose SE, Walsh M, Griffin M, Janke AL, McMahon KL, et al. Correlation of quantitative EEG in acute ischemic stroke with 30-day NIHSS score comparison with diffusion and perfusion MRI. Stroke. 2004;35(4):899-903.
101. Morelli M, Carta AR, Kachroo A, Schwarzschild MA. Pathophysiological roles for purines: adenosine, caffeine and urate. Progress in brain research. 2010;183:183-208.
102. Cui Q, Pitt JE, Pamukcu A, Poulin J-F, Mabrouk OS, Fiske MP, et al. Blunted mGluR Activation Disinhibits Striatopallidal Transmission in Parkinsonian Mice. Cell Reports. 2016;17(9):2431-44.
103. Soikkeli R, Partanen J, Soininen H, Pääkkönen A, Riekkinen P. Slowing of EEG in Parkinson's disease. Electroencephalography and clinical neurophysiology. 1991;79(3):159-65.
104. Neufeld M, Blumen S, Aitkin I, Parmet Y, Korczyn A. EEG frequency analysis in demented and nondemented parkinsonian patients. Dementia and Geriatric Cognitive Disorders. 1994;5(1):23-8.
105. Cozac VV, Gschwandtner U, Hatz F, Hardmeier M, Rüegg S, Fuhr P. Quantitative EEG and Cognitive Decline in Parkinson's Disease. Parkinson's Disease. 2016;2016.
106. Miyazaki I, Asanuma M. Serotonin 1A Receptors on Astrocytes as a Potential Target for the Treatment of Parkinson's Disease. Current medicinal chemistry. 2016;23(7):686-700.
107. Rappold PM, Tieu K. Astrocytes and therapeutics for Parkinson's disease. Neurotherapeutics. 2010;7(4):413-23.